\documentclass[
aps,%
11pt,%
final,
notitlepage,%
oneside,%
onecolumn,%
nobibnotes,%
nofootinbib,%
preprintnumbers,%
superscriptaddress,%
showpacs,%
centertags,%
showkeys,%
amsmath,%
amssymb]{revtex4}

\usepackage{graphicx} 
\usepackage{dcolumn} 
\usepackage{feynmp}
\usepackage{multirow}

\begin{document}


\title{Single Top Quark Production at the Fermilab Tevatron Collider}
\author{Jovan Mitrevski}
\email{jmitrevs@scipp.ucsc.edu}
\affiliation{Santa Cruz Institute for Particle Physics, University of California, Santa Cruz, CA  95064, USA}
\collaboration{on behalf of the CDF and D0 Collaborations}
\noaffiliation

\begin{abstract}
First evidence for single top quark production has recently been found by both the D0 and CDF experiments. By combining three analyses, D0 measured a cross section of $\sigma\left(p\bar{p} \rightarrow tb+X,~tqb+X\right) = 4.7 \pm 1.3 \,{\rm pb}$, with a significance of 3.6 standard deviations, and CDF's matrix elements analysis measured a cross section of $\sigma\left(p\bar{p} \rightarrow tb+X,~tqb+X\right) = 3.0 ^{+1.2}_{-1.1}\,{\rm pb}$, with a significance of 3.1 standard deviations. These analyses also provided the first direct measurements of the CKM matrix element, $|V_{tb}|$, without assuming unitarity. This talk briefly describes the latest single top production results from the two experiments.
\end{abstract}


\maketitle

\section{Introduction}

\begin{figure}[tbhp]
\begin{center}
\vspace{0.2in}
\begin{minipage}{45mm}
\begin{center}
\begin{fmfgraph*}(40,22)
\fmfleft{db,u}\fmfright{bb,t}
\fmflabel{$u$}{u}
\fmflabel{$\bar{d}$}{db}
\fmflabel{$t$}{t}
\fmflabel{$\bar{b}$}{bb}
\fmf{fermion}{u,v1,db}
\fmf{boson,label=$W$,label.side=left}{v1,v2}
\fmf{fermion}{bb,v2,t}
\end{fmfgraph*}
\end{center}
\end{minipage}
\put(-50,0){(a)}
\hspace{2cm}
\begin{minipage}{45mm}
\begin{center}
\begin{fmfgraph*}(40,22)
\fmfleft{b,u}\fmfright{t,d}
\fmflabel{$u$}{u}
\fmflabel{$d$}{d}
\fmflabel{$t$}{t}
\fmflabel{$b$}{b}
\fmf{fermion}{b,v1,t}
\fmf{boson,label=$W$,label.side=right}{v1,v2}
\fmf{fermion}{u,v2,d}
\end{fmfgraph*}
\end{center}
\end{minipage}
\put(-50,0){(b)}
\vspace{0.1in}
\caption[LO Feynman diagrams for single top quark production via the $s$-channel and $t$-channel processes]{LO Feynman diagrams for single top production via the (a) $s$-channel and (b) $t$-channel  process.}
\label{fig:feynman}
\end{center}
\end{figure}
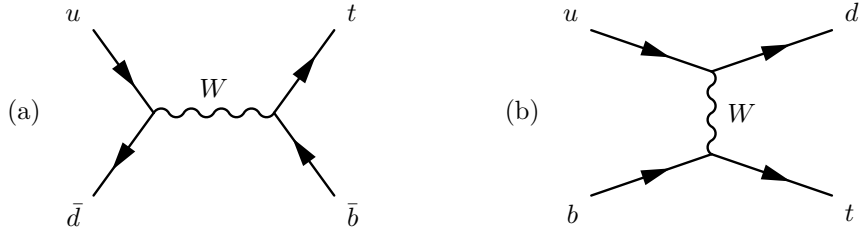

The top quark was discovered in 1995 by the D0 and CDF collaborations, produced in pairs by the strong force~\cite{Abachi:1995iq, Abe:1995hr}. However, the top quark can also be produced singly, by the electroweak force, providing a powerful probe of charged-current interactions involving the top quark. The two categories of single top processes that are significant at the Fermilab Tevatron Collider are $s$-channel, which involve the exchange of a time-like $W$ boson, and $t$-channel, which involve the exchange of a space-like $W$ boson. Representative leading order (LO) Feynman diagrams are given in Fig.~\ref{fig:feynman}. For proton-antiproton collisions with $\sqrt{s} = 1.96\,\mathrm{TeV}$, using a top quark mass of $175\,\mathrm{GeV}$, the theoretical cross section at next to leading order (NLO) for $s$-channel production is 
$\sigma_s = 0.88\pm0.11\,\mathrm{pb}$,
and for $t$-channel production, 
$\sigma_t= 1.98\pm0.25\,\mathrm{pb}$~\cite{Sullivan:2004ie}. 
The given cross sections are for the sum of top and antitop production. In addition to $s$-channel and $t$-channel production, a single top quark can also be created in association with an an on-shell $W$ boson, but this process is negligible at the Fermilab Tevatron Collider.
\begin{fmffile}{singletop-lo}
\end{fmffile}

One of the main reasons to measure the single top production cross section is because it is proportional to $|V_{tb}|^2$; thus measuring it is a way to directly measure the CKM matrix element $|V_{tb}|$ without assuming three generations or unitarity. Additionally, single top quark production is sensitive to physics beyond the standard model~\cite{Tait:2000sh}. Lastly, single top production is a background to Higgs and other beyond the standard model searches.

This talk summarizes the latest single top quark searches performed by the D0 and CDF experiments. The D0 results are based the analysis published in Ref.~\cite{singletop-prl}, using $0.9\,\mathrm{fb}^{-1}$ of data, with some updates to the techniques. The CDF results are based on $1.5\,\mathrm{fb}^{-1}$ of data.

\section{Event Selection}

According to the standard model, the top quark decays almost exclusively into a $b$-quark and a $W$ boson. Figure~\ref{fig:feynman} shows that the $s$-channel single top quark process has a second $b$-quark in the final state while the $t$-channel single top quark process has a light \emph{spectator} quark. 
For $t$-channel processes, the initial state $b$-quark comes from gluon splitting, so sometimes another $b$-quark, generally with low transverse momentum, is found in the final state. That is why $t$-channel production is sometimes labeled $tqb$. Additional jets can come from radiated gluons. For events in which the $W$ boson decays into either an electron or a muon plus a neutrino, the main backgrounds are $W$+jets, including $W$+heavy flavor, $t\bar{t}$, and QCD multijet events where one jet is misidentified as a lepton or a lepton strays from a jet. Diboson and $Z$+jets events form a smaller background.

The selected events were required to have one reconstructed charged electron or muon, $\not\!\!E_\mathrm{T}$ from the neutrino, and at least two jets. CDF required exactly two jets, while D0 used events with two, three, and in some analyses four jets. At least one jet had to be identified as coming from a $b$-quark by lifetime tagging. Additional cuts were applied to improve the purity of the sample by reducing the fake lepton background, either by cutting on the difference in azimuthal angle between the $\not\!\!E_\mathrm{T}$ and the jets or leptons, or on the transverse mass of the reconstructed $W$ boson, or on the $\not\!\!E_\mathrm{T}$ significance. The events were triggered by a lepton+jets trigger at D0 or a single lepton trigger at CDF. The CDF triggers had a $\not\!\!E_\mathrm{T}$ requirement for high pseudorapidity electrons. Table~\ref{tab:yields} shows the resulting yields for the CDF and D0 analyses.
\renewcommand{\multirowsetup}{\centering}
\begin{table}[tb]
\caption{\label{tab:yields} Predicted event yield at CDF with $1.5\,\mathrm{fb}^{-1}$ and at D0 with $0.9\,\mathrm{fb}^{-1}$ for electrons and muons combined and at least one $b$-tagged jet. In the D0 analyses the diboson and $Z$+jets backgrounds are not separately modeled but are instead included in the $W$+jets normalization.}
\vspace{0.1in}
\begin{ruledtabular}
\begin{tabular}{c|cccc}
\multirow{2}{2cm}{Source} & CDF Event Yields & \multicolumn{3}{c}{D0 Event Yields} \\
   & 2 jets & 2 jets & 3 jets & 4 jets \\ \hline 
$s$-channel	 & $23.9 \pm 6.1$ & $16 \pm 3$ & $8 \pm 2$  & $2 \pm 1$ \\
$t$-channel	 & $37.0 \pm 5.4$ & $20 \pm 4$ & $12 \pm 3$ & $4 \pm 1$ \\ \hline
$t\bar{t} \rightarrow \ell \ell$ & \multirow{2}{2cm}{$85.3 \pm 17.8$} & $39 \pm 9$ & $32 \pm 7$ & $11 \pm 3$ \\
$t\bar{t} \rightarrow \ell$+jets &     & $20 \pm 5$ & $103 \pm 25$ & $143 \pm 33$ \\
Diboson   & $40.7 \pm 4.0$ & & & \\
$Z$+jets	 & $13.8 \pm 2.0$& & & \\
$W$+bottom & $ 319.6 \pm 112.3$ & $261 \pm 55$ & $120 \pm 24$ & $35 \pm 7$  \\
$W$+charm\footnote{only $W+c\bar{c}$ in D0} & $324.2 \pm 115.8$ & $151 \pm 31$ & $85 \pm 17$  & $23 \pm 5$ \\
$W$+light jets\footnote{contains $W+c$ in D0} & $214.6 \pm 27.3$ & $119 \pm 25$ & $43 \pm 9$ & $12 \pm 2$ \\
Multijets & $44.5 \pm 17.8$ & $95 \pm 19$ & $77 \pm 15$ & $29 \pm 6$ \\
Total background & $1042.8 \pm 218.2$ & $686 \pm 41$ & $460 \pm 39$ & $253 \pm 38$ \\
\hline
Observed & $1078$ & $697$ & $455$ & $246$ \\
\end{tabular}
\end{ruledtabular}
\end{table}
%



\section{Multivariate Analyses}

In order to extract the signal from the much bigger background, multivariate analyses were used. Such techniques generally build a discriminant of the type:
\begin{equation}
\label{eq:disc}
D(\mathbf{x}) = \frac{p(\mathbf{x}|S)}{p(\mathbf{x}|S) + p(\mathbf{x}|B)}  
\end{equation}
where $p(\mathbf{x}|S)$ is the probability density to observe the given configuration ($\mathbf{x}$) of jets and leptons given that the event is signal, and $p(\mathbf{x}|B)$ is the corresponding probability density given that the event is background.

Both D0 and CDF performed a single top search using the matrix elements technique, which calculates $p(\mathbf{x}|S)$ and $p(\mathbf{x}|B)$ using 
\begin{equation}
p(\mathbf{x}|\mathrm{process}_i) = \frac{1}{\sigma_i}\frac{d\sigma_i}{d\mathbf{x}},
\end{equation}
where $\sigma_i$ is the cross section of a given process, with numerical (usually Monte Carlo) integration. Both used MadGraph LO matrix elements for the calculation~\cite{Maltoni:2002qb}. One complication is that the event configuration, $\mathbf{x}$, refers to reconstructed objects, such as jets, while the matrix element uses the parton-level event configuration, represented by $\mathbf{y}$. The parton-level values are summed over:
\begin{equation}
\label{dsigma}
\frac{d\sigma}{d\mathbf{x}} = \sum_j \int d\mathbf{y}
\left[ f_{1, j}(\mathbf{y}, Q^{2})\, f_{2, j}(\mathbf{y}, Q^{2})\,
\frac{d\sigma_{hs,j}}{d\mathbf{y}}(\mathbf{y})\,W(\mathbf{x}|\mathbf{y},j)\,
\Theta_{\mathrm{parton}}(\mathbf{y}) \right],
\end{equation}
where $d\sigma_{hs}/d\mathbf{y}$ is the differential cross section for the hard scatter collision, $f$ represent the parton distribution functions for the proton and antiproton, $W$ the transfer function relating the probability to observe event configuration $\mathbf{x}$ given event configuration $\mathbf{y}$, and $\Theta$ represents parton-level cuts. The summation is over the various ways to assign jets to partons.

\begin{figure}[tb]
\begin{center}
\includegraphics[width=0.4\textwidth]{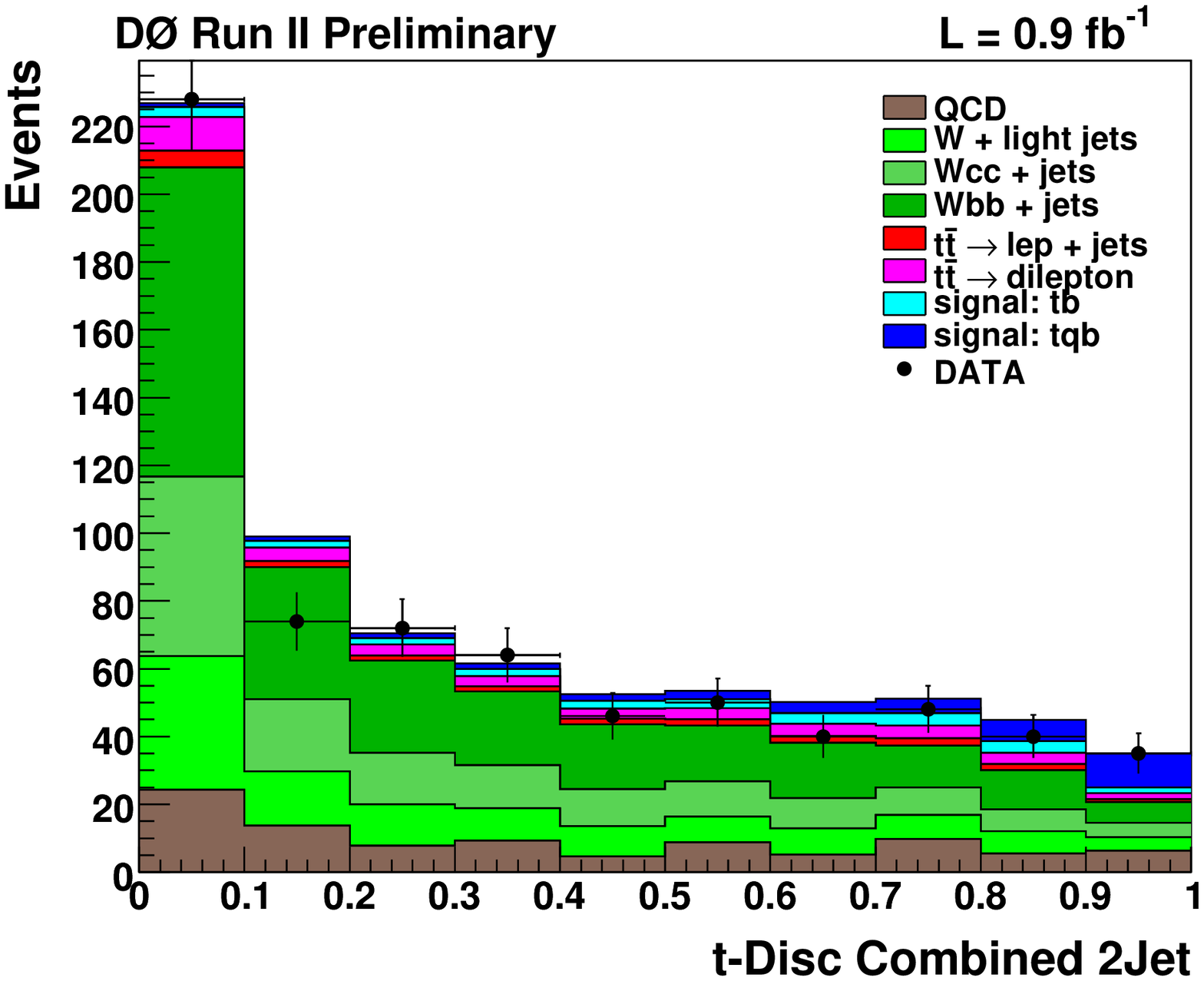}
\put(-58,0){(a)}
\hspace{1cm}
\includegraphics[width=0.42\textwidth]{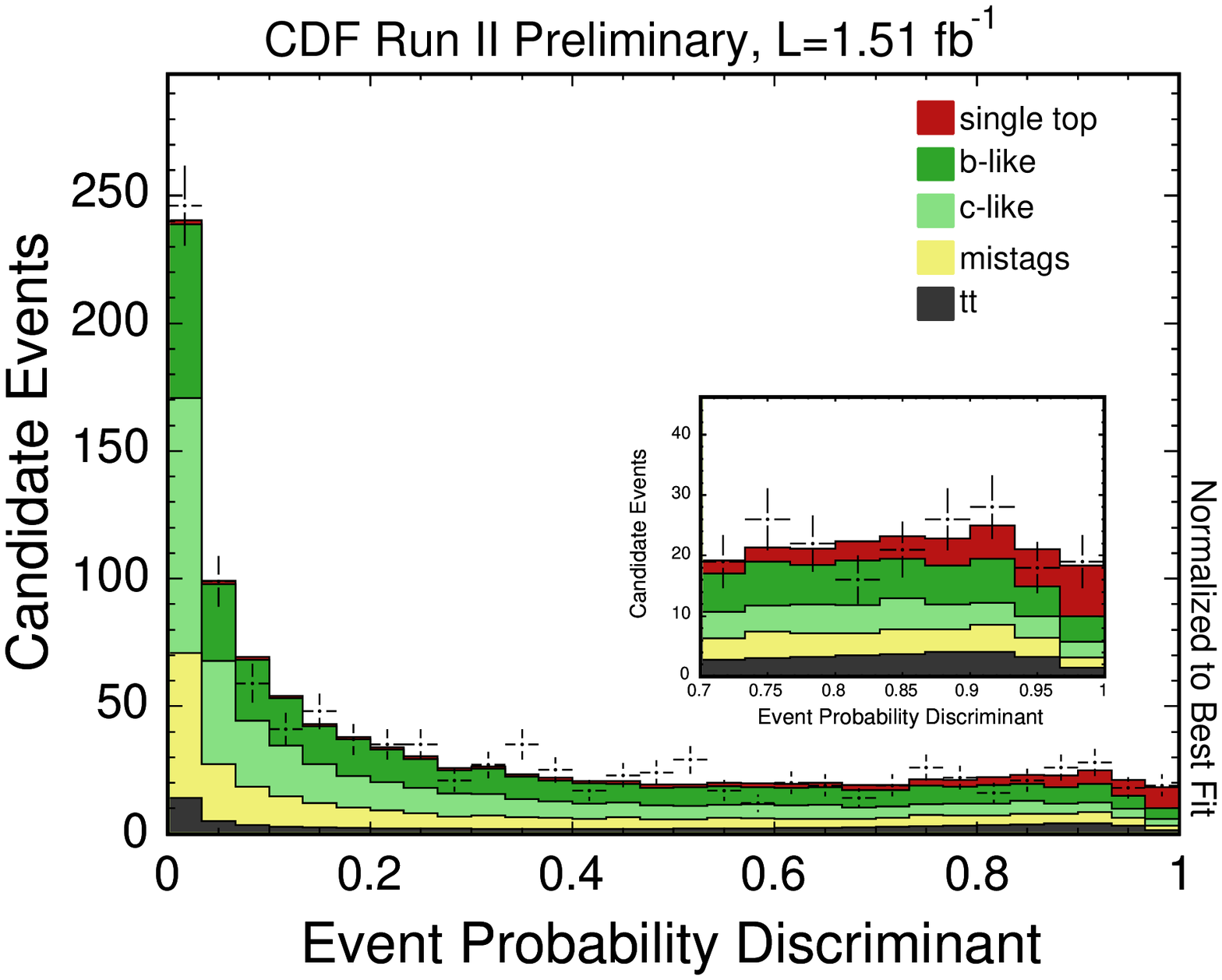}
\put(-60,0){(b)}
\caption[ME Discriminant output]{Discriminant output of the ME analyses for D0 (a) and CDF (b).}
\label{fig:medisc}
\end{center}
\end{figure}

D0's current matrix elements results are an update to those published in Ref.~\cite{singletop-prl}, with the primary change being the addition of a $t\bar{t}$ matrix element to the three-jet discriminants. Nevertheless, the main sensitivity remains in the two-jet events. Figure.~\ref{fig:medisc}(a) shows the output for the $t$-channel discriminant for two-jet events used by D0, normalized to the measured cross section, and Fig~\ref{fig:medisc}(b) show the output of the combined discriminant from CDF's analysis, normalized similarly.


D0 performed an analysis using boosted decision trees, which was the main analysis described in Ref.~\cite{singletop-prl}. It has not been updated, but it nevertheless remains the one with the greatest measured significance. Decision trees work by choosing a sequence of cuts, but unlike regular cuts-based analyses, they look on both sides of a cut, adding additional cuts until an end criteria is satisfied. The sequence of cuts forms a binary tree, and each leaf has a purity ($p$) associated with it, describing what fraction of the leaf content is signal. For a given event, $D(\mathbf{x})$ from Eq.~\ref{eq:disc} is approximated as $D(\mathbf{x}) = p_\mathbf{x}$, where $p_\mathbf{x}$ is the purity of the leaf where the event is assigned. Boosting improves the discriminating power by averaging over a number of trees, each tree being modified in such a way as to improve the classification of \emph{misclassified} events, which are signal events that fall in a background-like leaves or background events that fall in a signal-like leaves. Such events are given a boost in their weight, thereby making the tree-growing algorithm give them preference.

D0 used 49 variables as inputs to the decision trees, which can be classified in three categories: individual object kinematics, global event kinematics, and angular correlations. Some variables with high discriminating power are the invariant mass of all the jets in the event, the invariant reconstructed top quark mass, the angle between the highest-$p_\mathrm{T}$ $b$-tagged jet and the lepton in the rest frame of the reconstructed top quark, and the lepton charge times the pseudorapidity of the untagged jet ($Q\times \eta$). 



CDF performed an analysis using a likelihood discriminant. For the $t$-channel discriminant, it used the scalar sum of the jets, charged lepton, and $\not\!\!E_\mathrm{T}$ ($H_\mathrm{T}$); the cosine of the angle between the lepton and the spectator jet; $Q\times \eta$; the dijet invariant mass; the log of the $t$-channel matrix element; the output of a neural network jet flavor separator; and the $\chi^2$ of the kinematic solver, which reconstructs the top quark and the $W$~boson. The $s$-channel discriminant was built similarly. 


D0 performed an analysis using a Bayesian neural network. This analysis has been updated from what was published in Ref.~\cite{singletop-prl}, with much improved significance. Unlike a regular neural network which uses one set of network parameters ($\mathrm{w}_0$) determined by the training process to approximate $D(\mathrm{x}) = n(\mathrm{x}, \mathrm{w}_0)$, a Bayesian neural network integrates over all possible weights:
\begin{equation}
\label{eq:BNN}
n(\mathbf{x}) = \int n(\mathbf{x}, \mathbf{w}) \, p(\mathbf{w}|T) \, d\mathbf{w},
\end{equation}
where $p(\mathbf{w}|T)$ represents the probability density to have training weights $\mathbf{w}$ given training data $T$. Practically, the algorithm is implemented by averaging over a number of individual neural networks, so the Bayesian aspect can be viewed as an analogue to the boosting used for the decision trees.

The Bayesian neural network analysis used between 18 and 25 variables in each analysis channel. As an example, the variables with highest discriminating power in the electron, two jet, one $b$-tagged jet channel are the invariant mass of the leading two jets, the invariant reconstructed top quark mass, $Q\times \eta$, and the $p_{T}$ of the leading two jets summed vectorially. 

\section{Results}

The measured cross sections for the three analyses performed by D0, along with their combination, and the two analyses performed by CDF, are shown in Table~\ref{tab:xsec}. The combination was done using the Best Linear Unbiased Estimate (BLUE) method~\cite{Lyons:1988rp}. 
\begin{table}[tb]
\caption{\label{tab:xsec} The measured cross sections. The $s$-channel + $t$-channel cross section measurements assume a standard model cross section ratio of $\sigma_s/\sigma_t = 0.44$.}
\vspace{0.1in}
\begin{ruledtabular}
\begin{tabular}{c|ccc}
\multirow{2}{2cm}{Analysis} & \multicolumn{3}{c}{Cross Section (pb)}\\
         &  $s$-channel + $t$-channel & $s$-channel & $t$-channel \\ \hline
Boosted decision trees & $4.9 \pm 1.4$ & $1.0 \pm 0.9$ & $4.2^{+1.8}_{-1.4}$ \\
Matrix elements (D0) & $4.8^{+1.6}_{-1.4}$ & & \\
Bayesian NN & $4.4^{+1.6}_{-1.4}$ & & \\
D0 Combined & $4.7 \pm 1.3$ & & \\ \hline
Matrix elements (CDF)  & $3.0^{+1.2}_{-1.1}$ & $1.1^{+1.0}_{-0.8}$ & $1.9^{+1.0}_{-0.9}$ \\
Likelihood  & $2.7^{+1.3}_{-1.1}$ & $1.1^{+1.4}_{-1.1}$ & $1.3^{+1.2}_{-1.0}$ \\
\end{tabular}
\end{ruledtabular}
\end{table}
The boosted decision trees analysis, and both analyses at CDF also measured separate $s$-channel and $t$-channel cross sections, which are also given in the table. The expected and measured significances of the analyses are given in Table~\ref{tab:sig}.
\begin{table}[tb]
\caption{\label{tab:sig} The expected and measured significances for the various analyses.}
\vspace{0.1in}
\begin{ruledtabular}
\begin{tabular}{c|cc|cc}
\multirow{2}{2cm}{Analysis} & \multicolumn{2}{c|}{Expected} & \multicolumn{2}{c}{Measured} \\
& p-value & Significance (std. dev.) & p-value & Significance (std. dev.) \\ \hline
Boosted decision trees & 0.0177 & 2.1 & 0.00037 & 3.4 \\
Matrix elements (D0) & 0.0307 & 1.9 & 0.00082 & 3.2 \\
Bayesian NN & 0.0155 & 2.2 & 0.00083 & 3.1 \\
D0 Combined & 0.0105 & 2.3 & 0.00014 & 3.6 \\
\hline
Matrix elements (CDF)  & 0.0013 & 3.0 & 0.0009 & 3.1 \\
Likelihood  & 0.0020 & 2.9 & 0.0031 & 2.7 \\
\end{tabular}
\end{ruledtabular}
\end{table}

Both D0 and CDF measured $\left| V_{tb} \right|$ within the single top analyses.  These measurements make no assumptions on the unitarity of CKM or the number of families, but they do require a few assumptions. The first assumption is that the observed single top quarks were produced an interaction with a $W$ boson. The second assumption is that $|V_{tb}|^2 \gg |V_{td}|^2 + |V_{ts}|^2$, which is experimentally supported by the $\mathrm{B}(t \rightarrow Wb) / \mathrm{B}(t\rightarrow Wq)$ measurements done on $t\bar{t}$ events~\cite{Acosta:2005hr,Abazov:2006bh}. A third assumption is that the $Wtb$ vertex is CP-conserving and of the V$-$A form, though possibly of anomalous strength, which is parametrized by a multiplicative constant, $f^L_1$, equal to one in the standard model. Additional theoretical uncertainties were added to the measurements of $|V_{tb}|$~\cite{Sullivan:2004ie}.

On D0 the $\left| V_{tb} \right|$ measurement was done in the scope of the decision trees analysis and was published in Ref.~\cite{singletop-prl}. The final results are:
\begin{itemize}
\item $\left|V_{tb}f^L_1\right| = 1.31^{+0.25}_{-0.21}$, if $\left|V_{tb}f^L_1\right|$ is not constrained to be between zero and one,
\item $\left|V_{tb}\right| > 0.68$ at 95\% C.L., assuming that $\left|V_{tb}\right|$ is between zero and one and $f^L_1 = 1$.
\end{itemize}

On CDF, both the matrix elements analysis and the likelihood analysis performed $\left|V_{tb}\right|$ measurements. The results without being constrained to being between zero and one are:
\begin{itemize}
\item $\left|V_{tb}f_1^L\right| = 1.02 \pm 0.18 \mathrm{(expt)} \pm 0.07\mathrm{(theory)}$ for the matrix elements analysis,
\item $\left|V_{tb}f_1^L\right| = 0.97^{+0.21}_{-0.19} \mathrm{(expt)} \pm 0.07\mathrm{(theory)}$ for the likelihood analysis.
\end{itemize}
Assuming that $\left|V_{tb}\right|$ is between zero and one and $f^L_1 = 1$,
\begin{itemize}
\item $\left|V_{tb}\right| > 0.55$ at 95\% C.L for the matrix elements analysis,
\item $\left|V_{tb}\right| > 0.52$ at 95\% C.L for the likelihood analysis.
\end{itemize}

\section{Summary and Conclusion}

D0 and CDF have both found evidence for single top production. Combining the three analyses at D0 resulted in a cross section measurement of $\sigma\left(p\bar{p} \rightarrow tb+X,~tqb+X\right) = 4.7 \pm 1.3 \,{\rm pb}$, with a significance of 3.6 standard deviations. With the matrix elements analysis, CDF measured a cross section of $\sigma\left(p\bar{p} \rightarrow tb+X,~tqb+X\right) = 3.0 ^{+1.2}_{-1.1}\,{\rm pb}$, with a significance of 3.1 standard deviations. The first direct measurements of $|V_{tb}|$ were performed without assuming three generational unitarity, limiting $\left|V_{tb}\right| > 0.68$ at 95\% C.L using the D0 decision tree analysis, and $\left|V_{tb}\right| > 0.55$ at 95\% C.L using the CDF matrix elements analysis.


\bibliography{hs07_singletop}

\end{document}